\documentclass[]{article}
\usepackage{latexsym}
\usepackage{amsmath}
\usepackage{amssymb}
\usepackage{amsfonts}
\usepackage{graphicx}
\usepackage{epsfig}
\usepackage{array}
\usepackage[all]{xy}
 
\newtheorem{theorem}{Theorem}

\newtheorem{ExampleDef}{Example}[section]

\newcommand{\Example}[3]{
  \begin{list}{}{
      \setlength{\leftmargin}{1em}}     
    \item                               
    \small                              
    \begin{ExampleDef} \rm              
      {\bf \hspace{-1ex}: #1}           
      #2                                
      \hfill {\large \boldmath $\Box$}  
      \label{ex:#3}                      
    \end{ExampleDef}
  \end{list}}

\begin{document}
\begin{center}
{\Large {\bf Contraction Analysis of Time-Delayed Communications 
Using Simplified Wave Variables\par}}
\vspace{1.0em}
{\large Wei Wang and Jean-Jacques E. Slotine \par} 
{Nonlinear Systems Laboratory \\
Massachusetts Institute of Technology \\
Cambridge, Massachusetts, 02139, USA 
\\ wangwei@mit.edu, \ jjs@mit.edu \par}
\vspace{2em}
\end{center}

\begin{abstract}
We study stability of interacting nonlinear systems with time-delayed
communications, using contraction theory and a simplified wave
variable design inspired by robotic teleoperation. We show that
contraction is preserved through specific time-delayed feedback
communications, and that this property is independent of the values of
the delays. The approach is then applied to group cooperation with
linear protocols, where it shown that synchronization can be made
robust to arbitrary time delays.
\end{abstract}

\section{Introduction}
In many engineering applications, communications delays between
subsystems cannot be neglected.  Such an example is bilateral
teleoperation, where signals can experience significant transmission
delays between local and remote sites. Throughout the last decade,
both internet and wireless technologies have vastly extended practical
communication distances. Information exchange and cooperation can now
occur in very widely distributed systems, making the effect of time
delays even more central.

In the context of telerobotics, \cite{spong89} proposed a control
law for force-reflecting teleoperators which preserves passivity, and
thus overcomes the instability caused by time delays. The idea was
reformulated in~\cite{niemeyer91} in terms of scattering or
``wave'' variables. Transmission of wave variables across communication
channels ensures stability without knowledge of the time
delay. Further extensions to internet-based applications were
developed~\cite{niemeyer01,niemeyer98,spong03}, in which
communication delays are variable.

Recently, \cite{winni00,jjs01} extended the application of wave
variables to a more general context by performing a nonlinear
contraction analysis~\cite{winni98,winni00,rouchon,egeland} of the
effect of time-delayed communications between contracting systems.
This paper modifies the design of the wave variables proposed
in~\cite{winni00,jjs01}. Specifically, a simplified form provides an
effective analysis tool for interacting nonlinear systems with
time-delayed feedback communications. For appropriate coupling terms,
contraction as a generalized stability property is preserved
regardless of the delay values.  The result also sheds a new light on
the well-known fact in bilateral teleoperation, that even small
time-delays in feedback PD controllers may create stability problems
for simple coupled second-order systems, which motivated approaches
based on passivity and wave variables~\cite{spong89,niemeyer91}. The
approach is then applied to study the group cooperation problem with
delayed communications.  We show that synchronization with linear
protocols~\cite{olfati04,moreau} is robust to time delays and
network connectivity, without requiring the delays to be known or
equal in all links. In a leaderless network, all the coupled elements
tend to reach a common state which varies according to the initial
conditions and the time delays, while in a leader-followers network
the group agreement point is fixed by the leader. The approach is
suitable to study both continuous and discrete-time models.

After brief reviews of nonlinear contraction and wave variables,
Section $2$ uses simplified wave variable forms to analyze
time-delayed feedback communications. The same approach is then
applied in Section $3$ to study group cooperation. Concluding remarks
are offered in Section $4$.

\section{Contraction Analysis of Time-Delayed Communications}
Inspired by the use of passivity~\cite{spong89} and wave
variables~\cite{niemeyer91} in force-reflecting teleoperation,
\cite{winni00} performed a contraction analysis of the effect of
time-delayed communications. As we will discuss in this
section, the form of the transmitted variables in~\cite{winni00} can
be simplified and applied to analyze time-delayed feedback
communications.

\subsection{Background I: Contraction Theory} 
We first summarize briefly some basic definitions and main results of
nonlinear contraction theory~\cite{winni98, winnithesis}. Consider a
nonlinear system
\begin{equation} \label{eq:nonlinear_form}
\dot{{\bf x}} = {\bf f}({\bf x},t)
\end{equation}
with ${\bf x} \in \mathbb{R}^{m \times 1}$ and ${\bf f}({\bf x},t)$
continuously differentiable. For any virtual
displacement\footnote{Virtual displacements are differential
displacements at fixed time borrowed from mathematical physics and
optimization theory. Formally, if we view the position of the system
at time $t$ as a smooth function of the initial condition ${\bf x}_o$
and of time, $\ {\bf x} = {\bf x}({\bf x}_o ,t)\ $, then $\ \delta
{\bf x} = \frac{\partial {\bf x}}{\partial {\bf x}_o} \ d {{\bf x}_o}\
$.} $\delta {\bf x}$ between two neighboring trajectories, we have
$$
 \frac{d}{dt} ({\delta {\bf x}}^T {\delta {\bf x}}) = 
2 \ {\delta {\bf x}}^T {\delta \dot{{\bf x}}} =  
2 \ {\delta {\bf x}}^T \frac{\partial {\bf f}}{\partial {\bf x}} \ {\delta {\bf x}}
  \le 2\ \lambda_{max}\ {\delta {\bf x}}^T {\delta {\bf x}}
$$ 
where $\lambda_{max}({\bf x},t)$ is the largest eigenvalue of the
symmetric part of the Jacobian ${\bf J}=\frac{\partial {\bf
f}}{\partial {\bf x}}$. Hence, if $\lambda_{max}({\bf x},t)$ is
uniformly strictly negative, any infinitesimal length $\| \delta {\bf
x} \|$ converges exponentially to zero. By path integration at fixed
time, this implies that all the solutions converge exponentially to a
single trajectory, independently of the initial conditions.

More generally, considering a coordinate transformation
\begin{equation} \label{eq:contraction-transformation}
{\delta {\bf z}} = {\bf \Theta} \delta {\bf x}
\end{equation}
with square matrix ${\bf \Theta} ({\bf x},t)$ uniformly invertible,
we have
$$
\frac{d}{dt} ({\delta {\bf z}}^T {\delta {\bf z}}) = 
2 \ {\delta {\bf z}}^T {\delta \dot{\bf z}} = 
2 \ {\delta {\bf z}}^T \ ( \dot{{\bf \Theta}} + 
{\bf \Theta} \frac{\partial {\bf f}}{\partial {\bf x}} ) {\bf \Theta}^{-1} 
\ \delta {\bf z}
$$ 
so that exponential convergence of $\| \delta {\bf z} \|$ to zero is guaranteed
if the  {\em generalized Jacobian matrix}
\begin{equation} \label{eq:general-jacobian}
{\bf F}=(\dot{\bf \Theta} + 
        {\bf \Theta} \frac{\partial {\bf f}}{\partial {\bf x}} ) 
        {\bf \Theta}^{-1}
\end{equation}
is uniformly negative definite. Again, this implies in turn that all
the solutions of the original system (\ref{eq:nonlinear_form})
converge exponentially to a single trajectory, independently of the
initial conditions. Such a system is called {\it contracting}, with
{\it metric} ${\bf \Theta}^T {\bf \Theta}$.
 
In the sequel, we will also use {\em asymptotical contraction} to
refer to asymptotic convergence of any $\| \delta {\bf z} \|$ to zero,
which implies global asymptotic convergence to a single trajectory.

\subsection{Background II: Wave Variables}
\begin{figure}[h]
\begin{center}
\epsfig{figure=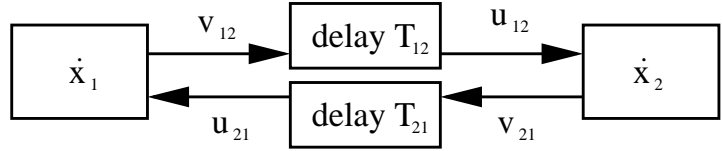,height=20mm,width=95mm}
\caption{Two interacting systems with delayed communications}
\label{fig:teleoperation}
\end{center}
\end{figure}

Wave variables~\cite{niemeyer91} are used in bilateral teleoperation
systems to guarantee the passivity~\cite{spong89} of time-delayed
transmissions. The idea was generalized in~\cite{winni00} by conducting
a contraction analysis on time-delayed transmission channels.
As illustrated in Figure~\ref{fig:teleoperation}, \cite{winni00}
considers two interacting systems of possibly different dimensions,
\begin{equation} \label{eq:two-delay-systems}
\begin{cases}
\ \dot{\bf x}_1\ =\  {\bf f}_1({\bf x}_1,t) + {\bf G}_{21} {\bf \tau}_{21} \\
\ \dot{\bf x}_2\ =\  {\bf f}_2({\bf x}_2,t) + {\bf G}_{12} {\bf \tau}_{12}
\end{cases}
\end{equation}
where ${\bf G}_{12}$, ${\bf G}_{21}$ are constant matrices and 
${\bf \tau}_{12}$, ${\bf \tau}_{21}$ have the same dimension. 
Communication between the two systems occurs by transmitting
intermediate ``wave'' variables, defined as
\begin{eqnarray*}
{\bf u}_{21} &=& {\bf G}_{21}^T {\bf x}_1 \ + \ {\bf \tau}_{21}  \ \ \ \ \ \ \ \ \ \ \ \ \ \ \ \ \
{\bf v}_{12} = {\bf G}_{21}^T {\bf x}_1 \ - \ {\bf \tau}_{21} \\ 
{\bf u}_{12} &=& {\bf G}_{12}^T {\bf x}_2 \ + \ {\bf \tau}_{12}  \ \ \ \ \ \ \ \ \ \ \ \ \ \ \ \ \ 
{\bf v}_{21} = {\bf G}_{12}^T {\bf x}_2 \ - \ {\bf \tau}_{12}
\end{eqnarray*}
Because of time-delays, one has
\begin{equation*}
{\bf u}_{12}(t) = {\bf v}_{12}(t-T_{12})  \ \ \ \ \ \ \ \ \ \ \ \ \ \ \ \ 
{\bf u}_{21}(t) = {\bf v}_{21}(t-T_{21})
\end{equation*}
where $T_{12}$ and $T_{21}$ are two positive constants. It
can be proved that, if both ${\bf f}_1$ and ${\bf f}_2$ are
contracting, the overall system is asymptotically contracting
independently of the time delays.

Note that subscripts containing two numbers indicate the communication
direction, e.g., subscript ``$12$'' refers to 
communication from node $1$ to $2$. This notation will be helpful
in Section~\ref{sec:group-delays}, where results will be extended
to groups of interacting subsystems.

\subsection{Time-Delayed Feedback Communications}
\label{sec:teleoperation}
Let us now consider modified choices of the transmitted
variables above. Specifically, for interacting
systems~(\ref{eq:two-delay-systems}), define now the transmitted
variables as
\begin{eqnarray} \label{eq:feedback-wave-variable}
{\bf u}_{21} &=& {\bf G}_{21}^T {\bf x}_1 \ + \ k_{21} {\bf \tau}_{21}
 \ \ \ \ \ \ \ \ \ \ \ \ \ \ \ \ \
{\bf v}_{12} = {\bf G}_{21}^T {\bf x}_1 \\
{\bf u}_{12} &=& {\bf G}_{12}^T {\bf x}_2 \ + \ k_{12} {\bf \tau}_{12}
  \ \ \ \ \ \ \ \ \ \ \ \ \ \ \ \ \
{\bf v}_{21} = {\bf G}_{12}^T {\bf x}_2 \nonumber
\end{eqnarray}
where $k_{12}$ and $k_{21}$ are two strictly positive constants.  
Consider, similarly to \cite{winni00,jjs01}, the differential length
$$
V\ =\ \frac{k_{21}}{2}\ \delta {\bf x}_1^T \delta {\bf x}_1\ +\    
    \frac{k_{12}}{2}\ \delta {\bf x}_2^T \delta {\bf x}_2\ +\ 
    \frac{1}{2}\ V_{1,2}
$$
where
\begin{equation*}
V_{1,2} = \int_{t-T_{12}}^t \delta {\bf v}_{12}^T \delta {\bf v}_{12}\ d\epsilon\ +\
        \int_{t-T_{21}}^t \delta {\bf v}_{21}^T \delta {\bf v}_{21}\ d\epsilon \\
\ - \ \int_{-T_{12}}^0 \delta {\bf v}_{12}^T \delta {\bf v}_{12}\ d\epsilon\ - \
        \int_{-T_{21}}^0 \delta {\bf v}_{21}^T \delta {\bf v}_{21}\ d\epsilon
\end{equation*}
Note that
\begin{eqnarray*}
V_{1,2} 
&=& \int_0^t (\ \delta {\bf v}_{12}^T \delta {\bf v}_{12}(\epsilon) -
                \delta {\bf v}_{12}^T \delta {\bf v}_{12}(\epsilon-T_{12}) +
                \delta {\bf v}_{21}^T \delta {\bf v}_{21}(\epsilon) -
                \delta {\bf v}_{21}^T \delta {\bf v}_{21}(\epsilon-T_{21}) \ )\ d\epsilon \\
&=&  \int_0^t (\ \delta {\bf v}_{12}^T \delta {\bf v}_{12}(\epsilon)\ -\
                 \delta {\bf u}_{12}^T \delta {\bf u}_{12}(\epsilon)\ +\
                 \delta {\bf v}_{21}^T \delta {\bf v}_{21}(\epsilon)\ -\
                 \delta {\bf u}_{21}^T \delta {\bf u}_{21}(\epsilon) \ )\ d\epsilon \\
&=& -2 \int_0^t (\ k_{21} \delta {\bf x}_1^T {\bf G}_{21} \delta {\bf \tau}_{21}  +
                   k_{12} \delta {\bf x}_2^T {\bf G}_{12} \delta {\bf \tau}_{12} )\ d\epsilon\ 
  - \int_0^t (\ k_{21}^2 \delta {\bf \tau}_{21}^T \delta {\bf \tau}_{21} +
                  k_{12}^2 \delta {\bf \tau}_{12}^T \delta {\bf \tau}_{12} )\ d\epsilon
\end{eqnarray*}
This yields
$$
\dot{V}\ =\ 
k_{21} \delta {\bf x}_1^T \frac{\partial {\bf f}_1}{\partial {\bf x}_1}\ \delta {\bf x}_1\ +\
k_{12} \delta {\bf x}_2^T \frac{\partial {\bf f}_2}{\partial {\bf x}_2}\ \delta {\bf x}_2\ -\ 
\frac{k_{21}^2}{2}\ \delta {\bf \tau}_{21}^T \delta {\bf \tau}_{21}\ -\
\frac{k_{12}^2}{2}\ \delta {\bf \tau}_{12}^T \delta {\bf \tau}_{12} 
$$
If ${\bf f}_1$ and ${\bf f}_2$ are both contracting with identity
metrics (i.e., if $\frac{\partial {\bf f}_1}{\partial {\bf x}_1}$ and
$\frac{\partial {\bf f}_2}{\partial {\bf x}_2}$ are both uniformly
negative definite), then $\ \dot{V} \le 0$, and $V$ is bounded and
tends to a limit.  Applying Barbalat's lemma~\cite{jjsbook} in
turn shows that, if $\ddot{V}$ is bounded, then $\dot{V}$ tends to
zero asymptotically, which implies that $\delta {\bf x}_1$, $\delta
{\bf x}_2$, $\delta {\bf \tau}_{12}$ and $\delta {\bf \tau}_{21}$ all
tend to zero.  Regardless of the values of the delays, all solutions
of system~(\ref{eq:two-delay-systems}) converge to a single
trajectory, independently of the initial conditions. In the sequel we
shall assume that $\ddot{V}$ can indeed be bounded as a consequence of
the boundedness of $V$.

This result has a useful interpretation. Expanding system
dynamics~(\ref{eq:two-delay-systems}) using
(\ref{eq:feedback-wave-variable}) yields
\begin{equation*}
\begin{cases}
\ \dot{\bf x}_1\ =\  {\bf f}_1({\bf x}_1,t) \ +\ \frac{1}{k_{21}}\ {\bf G}_{21}\ 
(\ {\bf G}_{12}^T {\bf x}_2(t-T_{21}) - {\bf G}_{21}^T {\bf x}_1(t)\ ) \\
\ \dot{\bf x}_2\ =\  {\bf f}_2({\bf x}_2,t) \ +\ \frac{1}{k_{12}}\ {\bf G}_{12}\ 
(\ {\bf G}_{21}^T {\bf x}_1(t-T_{12}) - {\bf G}_{12}^T {\bf x}_2(t)\ )
\end{cases}
\end{equation*}
so that the above communications in fact correspond to simple feedback
interactions.  If we assume further that ${\bf x}_1$ and ${\bf x}_2$
have the same dimension, and choose ${\bf G}_{12} = {\bf G}_{21} =
{\bf G}$, the whole system is actually equivalent to two
diffusively coupled subsystems
\begin{equation} \label{eq:two-delayed-diffusion}
\begin{cases}
\ \dot{\bf x}_1\ =\  {\bf f}_1({\bf x}_1,t) \ +\ \frac{1}{k_{21}}\ {\bf G} {\bf G}^T\ 
(\ {\bf x}_2(t-T_{21}) - {\bf x}_1(t)\ ) \\
\ \dot{\bf x}_2\ =\  {\bf f}_2({\bf x}_2,t) \ +\ \frac{1}{k_{12}}\ {\bf G} {\bf G}^T\ 
(\ {\bf x}_1(t-T_{12}) - {\bf x}_2(t)\ ) 
\end{cases}
\end{equation}
Note that constants $k_{12}$ and $k_{21}$ and thus coupling gains
along different diffusion directions can be different. We can thus
state
\begin{theorem} \label{th:two-delayed-diffusion}
Consider two subsystems, both contracting with identity metrics, and
interacting through time-delayed diffusion-like
couplings~(\ref{eq:two-delayed-diffusion}). Then the overall system is
asymptotically contracting.
\end{theorem}

\begin{figure}[h]
\begin{center}
\epsfig{figure=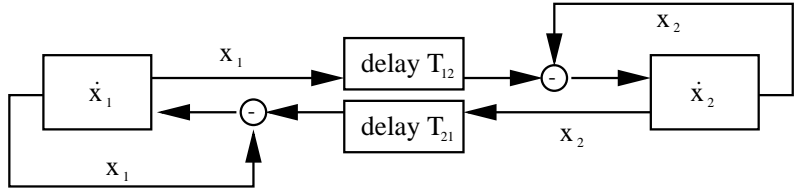,height=25mm,width=105mm}
\caption{Two interacting systems with time-delayed diffusion couplings}
\label{fig:delay-diffusion}
\end{center}
\end{figure}

{\em Remarks}

\begin{itemize}

\item Theorem~\ref{th:two-delayed-diffusion} does not contradict the
well-known fact in teleoperation, that even small time-delays in
bilateral PD controllers may create stability problems for coupled
second-order systems~\cite{spong89,niemeyer91,niemeyer98,spong03},
which motivates approaches based on passivity and wave variables.  In
fact, a key condition for contraction to be preserved is that the
diffusion-like coupling gains be symmetric positive semi-definite {\it
in the same metric} as the subsystems, as we shall illustrate later
in this section.

\item If there are no delays, i.e., $T_{12}=T_{21}=0$, 
then $V_{1,2}=0$. Thus, the analysis of the differential length 
yields that 
\begin{eqnarray*}
\dot{V} &=& \frac{1}{2} \frac{d}{dt} 
(\  k_{21} \delta {\bf x}_1^T \delta {\bf x}_1\ +\    
    k_{12} \delta {\bf x}_2^T \delta {\bf x}_2\ ) 
\ \le \ k_{21} \delta {\bf x}_1^T 
             \frac{\partial {\bf f}_1}{\partial {\bf x}_1} \delta {\bf x}_1\ +\
      k_{12} \delta {\bf x}_2^T
             \frac{\partial {\bf f}_2}{\partial {\bf x}_2} \delta {\bf x}_2 \\
&\le& \lambda_{max}\ 
(\  k_{21} \delta {\bf x}_1^T \delta {\bf x}_1\ +\    
    k_{12} \delta {\bf x}_2^T \delta {\bf x}_2\ )
\end{eqnarray*}
where
$$
\lambda_{max}\ =\ \max (\ 
   \lambda_{max}(\frac{\partial {\bf f}_1}{\partial {\bf x}_1})_s,\
   \lambda_{max}(\frac{\partial {\bf f}_2}{\partial {\bf x}_2})_s\ )
$$
This implies that $\delta {\bf x}_1$ and $\delta {\bf x}_2$ tend to
zero exponentially for contracting dynamics ${\bf f}_1$ and
${\bf f}_2$, i.e., the overall system is exponentially contracting.

\item If ${\bf f}_1$ and ${\bf f}_2$ are both contracting and
time-invariant, all solutions of
system~(\ref{eq:two-delayed-diffusion}) converge to a unique
equilibrium point, whose value is independent of the time delays and
the initial conditions. Indeed, in a globally contracting autonomous
system, all trajectories converge to a unique equilibrium
point~\cite{winni98, jjs03}, which implies that if ${\bf f}_1$ and
${\bf f}_2$ are both contracting and time-invariant, an equilibrium
point must exist for system~(\ref{eq:two-delayed-diffusion}) when
$T_{12}=T_{21}=0$. In turn, this point remains an equilibrium point
for arbitrary non-zero $T_{12}$ and $T_{21}$.  Since the delayed
system~(\ref{eq:two-delayed-diffusion}) also preserves contraction,
all solutions will converge to this point independently of the initial
conditions and the explicit values of the delays.

\item Theorem~\ref{th:two-delayed-diffusion} can be extended directly
to study more general connections between groups, such as
bidirectional meshes or webs of arbitrary size, and parallel
unidirectional rings of arbitrary length, both of which will be
illustrated through Section~\ref{sec:group-delays}. Inputs to 
the overall system can be provided through any of the subgroup dynamics.
\end{itemize}

\begin{figure}[h]
\begin{center}
\epsfig{figure=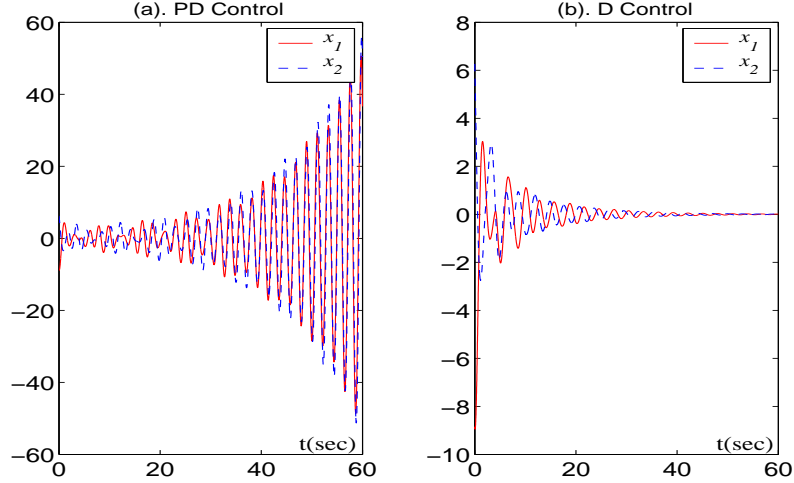,height=65mm,width=105mm}
\caption{Simulation results of two coupled mass-spring-damper
systems with (a) PD control and (b) D control. 
Parameters are $b=0.5$, $\omega^2 = 5$, $T_{12}=2s$,
$T_{21}=4s$, $k_d=1,$ $k_p=5$ in (a) and $k_p=0$ in (b). 
Initial conditions,
chosen randomly, are identical for the two plots.}
\label{fig:mass-spring-damping}
\end{center}
\end{figure}

\Example{}{Consider two identical second-order systems coupled through
time-delayed feedback PD controllers
\begin{equation*}
\begin{cases}
\ \ddot{x}_1 + b \dot{x}_1 + \omega^2 x_1\ =\
k_d (\dot{x}_2(t-T_{21})-\dot{x}_1(t)) + k_p (x_2(t-T_{21})-x_1(t)) \\
\ \ddot{x}_2 + b \dot{x}_2 + \omega^2 x_2\ =\ 
k_d (\dot{x}_1(t-T_{12})-\dot{x}_2(t)) + k_p (x_1(t-T_{12})-x_2(t))
\end{cases}
\end{equation*}
with $b>0$, $\omega > 0$.
If $T_{12}=T_{21}=0$, partial contraction analysis~\cite{wei03-2, wei03-1}
shows that $x_1$ and $x_2$
converge together exponentially regardless of initial conditions,
which makes the origin a stable equilibrium point.  If $T_{12}, T_{21}>0$,
a simple coordinate transformation yields
\begin{eqnarray*}
\left[ \begin{array}{l} \dot{x}_1 \\ \dot{y}_1 \end{array} \right] &=&
\left[ \begin{array}{c} \omega y_1 - b x_1  \\
                        -\omega x_1  \end{array} \right] +
{\bf K}
(\left[ \begin{array}{l} x_2(t-T_{21}) \\ y_2(t-T_{21}) \end{array} \right] -
 \left[ \begin{array}{l} x_1(t) \\ y_1(t) \end{array} \right]) \\
\left[ \begin{array}{l} \dot{x}_2 \\ \dot{y}_2 \end{array} \right] &=&
\left[ \begin{array}{c} \omega y_2 - b x_2  \\
                       -\omega x_2  \end{array} \right] +
{\bf K}
(\left[ \begin{array}{l} x_1(t-T_{12}) \\ y_1(t-T_{12}) \end{array} \right] -
 \left[ \begin{array}{l} x_2(t) \\ y_2(t) \end{array} \right])
\end{eqnarray*}
where 
$$
{\bf f}_1=\left[ \begin{array}{c} \omega y_1 - b x_1  \\
                                  -\omega x_1  \end{array} \right]\ \ \ \ \ \ \ \ \rm{and}\ \ \ \ \ \ \ {\bf f}_2=\left[ \begin{array}{c} \omega y_2 - b x_2  \\
                                  -\omega x_2  \end{array} \right]
$$ 
are both contracting with identity 
metric~\cite{wei03-1}. However, the transformed coupling gain
$$
{\bf K} =\left[ \begin{array}{ll} k_d & 0 \\ 
                 \frac{k_p}{\omega} & 0 \end{array} \right]
$$ 
is neither symmetric nor positive semi-definite for any $k_p \ne
0$. Contraction cannot be preserved in this case, and the coupled
systems turn out to be unstable for large enough delays as the
simulation result in
Figure~\ref{fig:mass-spring-damping}(a) illustrates.  In
Figure~\ref{fig:mass-spring-damping}(b) we set $k_p=0$ so that the
overall system is contracting according to
Theorem~\ref{th:two-delayed-diffusion}.
}{two-delayed-mechanical-systems}

The instability mechanism in the above example is actually very
similar to that of the classical Smale model~\cite{smale,wei03-1} of
spontaneous oscillation, in which two or more identical biological
cells, inert by themselves, tend to self-excited oscillations through
diffusion interactions. In both cases, the instability is caused by a
non-identity metric, which makes the transformed coupling gains lose
positive semi-definiteness.  Note that the relative simplicity with
which both phenomena can be analyzed makes fundamental use
of the notion of a metric, central to contraction theory.

\subsection{Other Simplified Forms of Wave-Variables}
Different simplifications of the original wave-variable design can be
made based on the same choice of $V$, yielding different qualitative
properties. For instance, the transmitted signals can be defined as
\begin{eqnarray*}
{\bf u}_{21} &=& {\bf G}_{21}^T {\bf x}_1 \ + \ k_{21} {\bf \tau}_{21} 
 \ \ \ \ \ \ \ \ \ \ \ \ \ \ \ \ 
{\bf v}_{12} = - \ k_{21} {\bf \tau}_{21}  \\ 
{\bf u}_{12} &=& {\bf G}_{12}^T {\bf x}_2 \ + \ k_{12} {\bf \tau}_{12} 
 \ \ \ \ \ \ \ \ \ \ \ \ \ \ \ \ 
{\bf v}_{21} = - \ k_{12} {\bf \tau}_{12}
\end{eqnarray*}
which leads to
\begin{eqnarray}\label{tau}
\dot{V} &=& 
k_{21} \delta {\bf x}_1^T \frac{\partial {\bf f}_1}{\partial {\bf x}_1} 
       \delta {\bf x}_1\ +\
k_{12} \delta {\bf x}_2^T \frac{\partial {\bf f}_2}{\partial {\bf x}_2} 
       \delta {\bf x}_2\ -\  
\frac{1}{2} \delta {\bf x}_1^T {\bf G}_{21} {\bf G}_{21}^T \delta {\bf x}_1\ -\ 
\frac{1}{2} \delta {\bf x}_2^T {\bf G}_{12} {\bf G}_{12}^T \delta {\bf x}_2 \nonumber \\
&=&
\delta {\bf x}_1^T(\ k_{21} \frac{\partial {\bf f}_1}{\partial {\bf x}_1}\ -
                     \frac{1}{2} {\bf G}_{21} {\bf G}_{21}^T\ ) \delta {\bf x}_1\ +\
\delta {\bf x}_2^T(\ k_{12} \frac{\partial {\bf f}_2}{\partial {\bf x}_2}\ -
                     \frac{1}{2} {\bf G}_{12} {\bf G}_{12}^T\ ) \delta {\bf x}_2 
\end{eqnarray}
and thus also preserves contraction through time-delayed communications.
Similarly to the previous section, 
if both ${\bf f}_1$ and ${\bf f}_2$ are contracting and time-invariant, the whole
system tends towards a unique equilibrium point, regardless of the delay values.
At this steady state,  
${\bf u}_{12}(\infty)\ =\ {\bf v}_{12}(\infty)$ and  
${\bf u}_{21}(\infty)\ =\ {\bf v}_{21}(\infty)$,
which immediately implies that
$$
{\bf G}_{21}^T \ {\bf x}_1(\infty)\ =\ {\bf G}_{12}^T \ {\bf x}_2(\infty)
$$ 
and, if ${\bf G}_{12} = {\bf G}_{21} = {\bf G}$ with ${\bf
G}$ of full rank, that 
$$
{\bf x}_1(\infty)\ =\ {\bf x}_2(\infty)
$$  
Thus, contrary to the case~(\ref{eq:feedback-wave-variable}) of diffusion-like 
couplings, the remote tracking ability of wave variables is preserved. 
\Example{}{Consider the example of two second-order systems 
\begin{equation*}
\begin{cases}
\ \ddot{x}_1 + b_1 \dot{x}_1 + \omega_1^2 x_1\ =\ F_{21}\ +\ F_e \\
\ \ddot{x}_2 + b_2 \dot{x}_2 + \omega_2^2 x_2\ =\ F_{12}
\end{cases}
\end{equation*}
where $F_e$ is an external force, and $F_{21}$, $F_{12}$ are internal forces
undergoing time delays. Performing a coordinate transformation, we get
new equations
\begin{eqnarray*}
\left[ \begin{array}{l} \dot{x}_1 \\ \dot{y}_1 \end{array} \right] &=&
\left[ \begin{array}{c} \omega_1 y_1 - b_1 x_1  \\
                   -\omega_1 x_1 + \frac{F_e}{\omega_1}  \end{array} \right]\ +\
 \left[ \begin{array}{l} 0 \\ \frac{F_{21}}{\omega_1} \end{array} \right] \\
\left[ \begin{array}{l} \dot{x}_2 \\ \dot{y}_2 \end{array} \right] &=&
\left[ \begin{array}{c} \omega_2 y_2 - b_2 x_2  \\
                       -\omega_2 x_2  \end{array} \right]\ +\
 \left[ \begin{array}{l} 0 \\ \frac{F_{12}}{\omega_2} \end{array} \right] 
\end{eqnarray*}
The signals being transmitted are defined as simplified wave variables
\begin{eqnarray*}
u_{21} &=& y_1 \ + \ \frac{k_{21}}{\omega_1} F_{21} 
 \ \ \ \ \ \ \ \ \ \ \ \ \ \ \ \
v_{12} = - \ \frac{k_{21}}{\omega_1} F_{21} \\ 
u_{12} &=& y_2 \ + \ \frac{k_{12}}{\omega_2} F_{12} 
 \ \ \ \ \ \ \ \ \ \ \ \ \ \ \ \
v_{21} = - \ \frac{k_{12}}{\omega_2} F_{12}
\end{eqnarray*}
Note that here ${\bf G}_{12} = {\bf G}_{21} =
\left[ \begin{array}{cc} 0 & 0 \\ 0 & 1 \end{array} \right]$, and that
although the variables $y_1$ and $y_2$ are virtual, their values
can be calculated based on $x_1$, $\dot{x}_1$ and
$x_2$, $\dot{x}_2$. According to the result we derived above,
the whole system will tend to reach an equilibrium point
asymptotically. This point is independent to the time delays and
satisfies $y_1(\infty)\ =\ y_2(\infty)$, i.e., 
$$
\frac{b_1}{\omega_1} x_1(\infty)\ =\ \frac{b_2}{\omega_2} x_2(\infty)
$$
}{tau_version}
\begin{figure}[h]
\begin{center}
\epsfig{figure=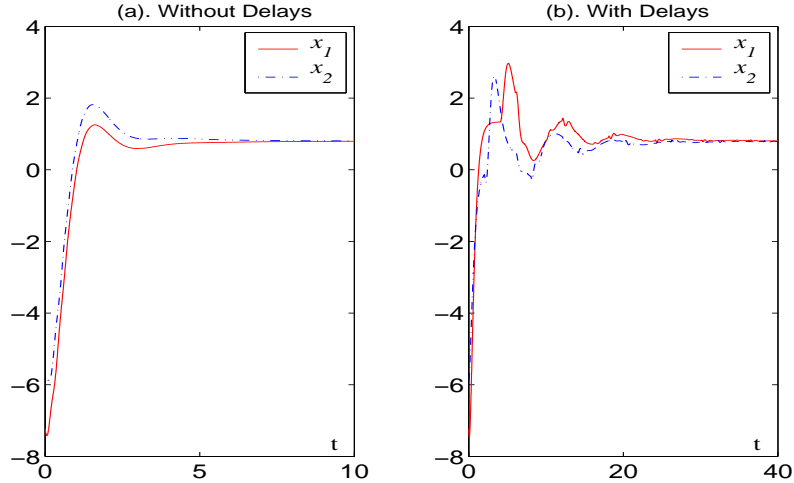,height=65mm,width=105mm}
\caption{Simulation results of Example~\ref{ex:tau_version} with
(a). $T_{12}=T_{21}=0$ and with (b). $T_{12}=2s$, $T_{21}=4s$.  The
parameters are $b_1=b_2=0.5$, $\omega_1^2=\omega_2^2=5$,
$k_{12}=k_{21}=0.2$, and $F_e=10$. Initial conditions, chosen
randomly, are identical for the two plots. Convergence to a common
equilibrium point independent to the time delays is achieved in both
cases.}
\label{fig:tau_version}
\end{center}
\end{figure}

Finally note that even if the subsystems are not contracting but have
upper bounded Jacobian, for instance, as limit-cycle oscillators,
the overall system still can be contracting
by choosing appropriate gains such that
\begin{equation*}
{\bf G}_{21} {\bf G}_{21}^T \ > \
   2 k_{21} ( \frac{\partial {\bf f}_1}{\partial {\bf x}_1} )_s
\ \ \ \ \ \ \ \ \ \rm{and} \ \ \ \ \ \ \ \ \ \ 
{\bf G}_{12} {\bf G}_{12}^T \ > \
   2 k_{12} ( \frac{\partial {\bf f}_2}{\partial {\bf x}_2} )_s
\end{equation*}
Here the transmission of wave variables performs a 
stabilizing role.

There are also other simplified forms of wave variables. For instance,
we can define the transmitted signals as
\begin{eqnarray*}
{\bf u}_{21} &=& {\bf G}_{21}^T {\bf x}_1 \ + \ k_{21} {\bf \tau}_{21} 
 \ \ \ \ \ \ \ \ \ \ \ \ \ \ \ \ \
{\bf v}_{12} = k_{21} {\bf \tau}_{21} \\
{\bf u}_{12} &=& {\bf G}_{12}^T {\bf x}_2 \ + \ k_{12} {\bf \tau}_{12}
 \ \ \ \ \ \ \ \ \ \ \ \ \ \ \ \ \
{\bf v}_{21} = k_{12} {\bf \tau}_{12}
\end{eqnarray*}
If both ${\bf f}_1$ and ${\bf f}_2$ are contracting and time-invariant, 
and if ${\bf G}_{12} = {\bf G}_{21} = {\bf G}$ with ${\bf G}$ of full rank, 
the whole system tends towards a unique equilibrium point such that
$ \ 
{\bf x}_1(\infty)\ =\ -\ {\bf x}_2(\infty)$.

\section{Group Cooperation with Time-Delayed Communications}
\label{sec:group-delays}
Recently, synchronization or group agreement has been the object of
extensive literature~\cite{kumar, jadbabaie03, leonard01, olfati_1,
pikovsky, reynolds, jjs03, strogatz03, tanner_2,
viscek95}. Understanding natural aggregate motions as in bird flocks,
fish schools, or animal herds may help achieve desired collective
behaviors in artificial multi-agent systems.  In our previous
work~\cite{wei03-2, wei03-1}, a synchronization condition was obtained
for a group of coupled nonlinear systems, where the number of the
elements can be arbitrary and the network structure can be very
general. In this section, we study a simplified continuous-time model
of schooling or flocking with time-delayed communications, and
generalize recent results in the
literature~\cite{olfati04,moreau}. In particular, we show that
synchronization is robust to time delays both for the leaderless case
and for the leader-followers case, without requiring the delays to be
known or equal in all links. Similar results are then derived for
discrete-time models.

\subsection{Leaderless Group} \label{sec:leaderless-delayed}
We first investigate a flocking model without group leader. The
dynamics of the $i$th element is given as
\begin{equation} \label{eq:linear_schooling}
\dot{\bf x}_i \ =\  
\sum_{j \in {\mathcal N}_i} {\bf K}_{ji}\ ({\bf x}_j - {\bf x}_i)  
 \ \ \ \ \ \ \ \ \ \ \ \ \ \ \ \ \ \ \ \ \ i = 1, \ldots, n
\end{equation}
where $\ {\bf x}_i \in \mathbb{R}^m\ $ denotes the states needed to reach 
agreements such as a vehicle's heading, attitude, velocity, etc. 
$\ {\mathcal N}_i\ $ denotes the set of the active neighbors of
element $i$, which for instance can be defined as the set of the nearest 
neighbors within a certain distance around $i$. $\ {\bf K}_{ji}\ $ is
the coupling gain, which is assumed to be symmetric and positive definite.
\begin{theorem} \label{th:linear_schooling}
Consider $n$ coupled elements with linear
protocol~(\ref{eq:linear_schooling}). The whole system will tend to 
reach a group agreement 
$$
{\bf x}_1(t) = \cdots = {\bf x}_n(t) = \frac{1}{n}
( {\bf x}_1(0) + \cdots + {\bf x}_n(0) )
$$ 
exponentially
if the network is connected, and the coupling links are either bidirectional
with ${\bf K}_{ji} = {\bf K}_{ij}$, or unidirectional but formed in
closed rings with identical gains.
\end{theorem}

Theorem~\ref{th:linear_schooling} is derived in~\cite{wei03-2,
wei03-1} based on partial contraction analysis, and the result can be
extended further to time-varying couplings (${\bf K}_{ji}={\bf
K}_{ji}(t)$), switching networks (${\mathcal N}_i={\mathcal
N}_i(t)$) and looser connectivity conditions.

Assume now that time delays are non-negligible in communications.
The dynamics of the $i$th element then turns to be
\begin{equation} \label{eq:linear_schooling_delayed}
\dot{\bf x}_i \ =\  
\sum_{j \in {\mathcal N}_i} {\bf K}_{ji}\ (\ {\bf x}_j(t-T_{ji}) - {\bf x}_i(t)\ )  
\end{equation}
\begin{theorem} \label{th:linear_schooling_delayed}
Consider $n$ coupled elements~(\ref{eq:linear_schooling_delayed}) with
time-delayed communications. Regardless of the explicit values of the delays,
the whole system will tend to reach a group agreement
${\bf x}_1(t) = \cdots = {\bf x}_n(t)$ asymptotically if
the network is connected, and the coupling links are either bidirectional
with ${\bf K}_{ji} = {\bf K}_{ij}$, or unidirectional but formed in
closed rings with identical gains.
\end{theorem}
{\bf Proof}: \ \ 
For notational simplicity, we first assume that all the links are bidirectional
with ${\bf K}_{ji} = {\bf K}_{ij}$, but the time delays could be different
along the opposite directions, i.e., $T_{ji} \ne T_{ij}$. Thus,
Equation~(\ref{eq:linear_schooling_delayed}) can be transformed to
$$
\dot{\bf x}_i\ =\ 
\sum_{j \in {\mathcal N}_i} {\bf G}_{ji} {\bf \tau}_{ji}
$$
where ${\bf \tau}_{ji}$ and correspondingly ${\bf \tau}_{ij}$ are defined
as  
\begin{eqnarray}  \label{eq:network-wave-variable}
{\bf u}_{ji} &=& {\bf G}_{ji}^T {\bf x}_i \ + \ {\bf \tau}_{ji} 
\ \ \ \ \ \ \ \ \ \ \ \ \ \ \
{\bf v}_{ij} = {\bf G}_{ji}^T {\bf x}_i \\
{\bf u}_{ij} &=& {\bf G}_{ij}^T {\bf x}_j \ + \ {\bf \tau}_{ij}
\ \ \ \ \ \ \ \ \ \ \ \ \ \ \ 
{\bf v}_{ji} = {\bf G}_{ij}^T {\bf x}_j \nonumber
\end{eqnarray}
with 
$\ {\bf G}_{ij} = {\bf G}_{ji}>0\ $ and
$\ {\bf K}_{ji} = {\bf K}_{ij} = {\bf G}_{ij} {\bf G}_{ij}^T\ $.
Define
\begin{equation} \label{eq:delay-network-lyapunov}
V\ =\ \frac{1}{2} \sum_{i=1}^n \delta {\bf x}_i^T \delta {\bf x}_i\ +\    
      \frac{1}{2} \sum_{(i,j) \in {\mathcal N}}\ V_{i,j}
\end{equation}
where ${\mathcal N}=\cup_{i=1}^n {\mathcal N}_i$ denotes the set of
all active links, and $V_{i,j}$ is defined as in
Section~\ref{sec:teleoperation} for each link connecting two nodes $i$
and $j$. Therefore
$$
\dot{V}\ =\ - \frac{1}{2} \sum_{(i,j) \in {\mathcal N}} 
           ( \delta {\bf \tau}_{ji}^T \delta {\bf \tau}_{ji} + 
             \delta {\bf \tau}_{ij}^T \delta {\bf \tau}_{ij} )
$$
One easily shows that $\ddot{V}$ is bounded. Thus
according to Barbalat's lemma, 
$\dot{V}$ will tend to zero asymptotically, which implies that,
$\forall (i,j) \in {\mathcal N}$, 
$\delta {\bf \tau}_{ji}$ and $\delta {\bf \tau}_{ij}$ tend to zero
asymptotically. Thus, we know that $\forall i$, $\delta \dot{\bf x}_i$
tends to zero. In standard calculus, a vanishing $\delta \dot{\bf x}_i$ 
does not necessarily mean that $\delta {\bf x}_i$ is convergent. But
it does in this case because otherwise it will conflict with the fact
that $\delta {\bf x}_i$ tends to be periodic with a constant period
$T_{ji}+T_{ij}$, which can be shown since
\begin{eqnarray} \label{eq:virtual-displayment-period}
\delta {\bf u}_{ji}(t) &=&  
{\bf G}_{ji}^T \delta {\bf x}_i(t) + \delta {\bf \tau}_{ji}(t)\ =\
{\bf G}_{ij}^T \delta {\bf x}_j (t-T_{ji}) \\
\delta {\bf u}_{ij}(t) &=&  
{\bf G}_{ij}^T \delta {\bf x}_j(t) + \delta {\bf \tau}_{ij}(t)\ =\
{\bf G}_{ji}^T \delta {\bf x}_i (t-T_{ij}) \nonumber
\end{eqnarray}
From~(\ref{eq:virtual-displayment-period}) we can also conclude that, if
$\forall i\ \delta {\bf x}_i$ is convergent, they 
will tend to a steady state
$$
\delta {\bf x}_1(t) = \cdots =  \delta {\bf x}_n(t) = {\bf c}
$$
where ${\bf c}$ is a constant vector whose value dependents on the 
specific trajectories we analyze. Moreover, we notice that, in the state-space, any point 
inside the region ${\bf x}_1 = \cdots = {\bf x}_n$ is invariant to 
(\ref{eq:linear_schooling_delayed}). By path integration, 
this implies immediately that, regardless of the delay values or the initial conditions, 
all solutions of the system~(\ref{eq:linear_schooling_delayed}) will tend to reach a 
group agreement ${\bf x}_1 = \cdots = {\bf x}_n$ asymptotically.

In case that coupling links are unidirectional but formed in
closed rings with coupling gains identical in each ring, we set
\begin{eqnarray*}
V &=& \frac{1}{2} \sum_{i=1}^n \delta {\bf x}_i^T \delta {\bf x}_i\ +\    
      \frac{1}{2} \sum_{(j \to i) \in {\mathcal N}} \ 
    (\ \int_{t-T_{ji}}^t \delta {\bf v}_{ji}^T \delta {\bf v}_{ji} d\epsilon -
       \int_{-T_{ji}}^0 \delta {\bf v}_{ji}^T \delta {\bf v}_{ji}\ d\epsilon\ ) \\
 &=& \frac{1}{2} \sum_{i=1}^n \delta {\bf x}_i^T \delta {\bf x}_i\ +\    
      \frac{1}{2} \sum_{(j \to i) \in {\mathcal N}} \ 
      \int_0^t (\ \delta {\bf v}_{ji}^T \delta {\bf v}_{ji}(\epsilon) -
                  \delta {\bf v}_{ji}^T \delta {\bf v}_{ji}(\epsilon-T_{ji})\ )\ d\epsilon
\end{eqnarray*}
and the rest of the proof is the same. The case when both types of links
are involved is similar. \hfill {\large \boldmath $\Box$}

\begin{figure}[h]
\begin{center}
\epsfig{figure=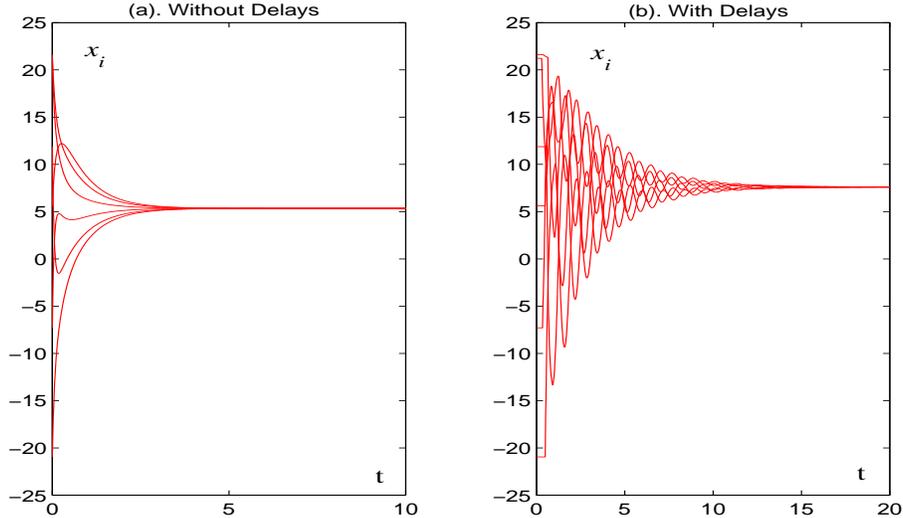,height=70mm,width=120mm}
\caption{Simulation results of Example~\ref{ex:flock-without-leader} without
delays and with delays. Initial conditions, chosen randomly, are
the same for each simulation. Group agreement is reached in both cases, although
the agreement value is different.}
\label{fig:delay-flock-without-leader}
\end{center}
\end{figure}

\Example{}{Compared to Theorem~\ref{th:linear_schooling}, the group
agreement point in Theorem~\ref{th:linear_schooling_delayed} generally
does not equal the average value of the initial conditions, but
depends on the values of the time delays.

Consider the cooperative group~(\ref{eq:linear_schooling_delayed}) with 
one-dimensional ${\bf x}_i$, $n=6$, and a two-way chain structure
$$ 
\xymatrix{ 1 \ar[r] & 2 \ar[l] \ar[r] & 3 \ar[l] \ar[r] & 4 \ar[l] \ar[r] 
         & 5 \ar[l] \ar[r] & 6 \ar[l] } 
$$
The coupling gains are set to be identical with value $k=5$. The delay values
are different, and each is chosen randomly around $0.5$ second.
Simulation results are plotted in Figure~\ref{fig:delay-flock-without-leader}.
}{flock-without-leader}

{\em Remarks}
\begin{itemize}
\item The conditions on coupling gains can be relaxed. For instance, if
the links are bidirectional, we do not have to ask ${\bf K}_{ij}={\bf K}_{ji}$.
Instead, the dynamics of the $i$th element could be
$$
\dot{\bf x}_i \ =\  
{\bf K}_i\ \sum_{j \in {\mathcal N}_i}\ (\ {\bf x}_j(t-T_{ji}) - {\bf x}_i(t)\ )  
$$ 
where ${\bf K}_i=\frac{1}{k_i} {\bf G}{\bf G}^T$ and ${\bf G}$ is
unique through the whole network. The proof is the same except that we
incorporate $k_i$ into the wave variables and the function $V$.  Such
a design brings more flexibility to cooperation-law design. The
discrete-time model studied in
Section~\ref{sec:discrete-flocking-delayed} in in this spirit. A
similar condition was derived in~\cite{chu} for a swarm model in the
no-delay case.

\item Model~(\ref{eq:linear_schooling}) with delayed communications
was also studied in~\cite{olfati04}, but the result is limited by the
assumptions that communication delays are equal in all links and that
the self-response part in each coupling uses the same time-delay. Recently,
\cite{moreau} independently analyzed the
system~(\ref{eq:linear_schooling_delayed}) in the scalar case with the
same assumption that delays are equal in all links.
\end{itemize}

\subsection{Leader-Followers Group}
Similar analysis can be applied to study coupled networks with group leaders.
Consider such a model
\begin{equation} \label{eq:linear_schooling_leader_delayed}
\dot{\bf x}_i =  
\sum_{j \in {\mathcal N}_i} {\bf K}_{ji}\ 
          ({\bf x}_j(t-T_{ji}) - {\bf x}_i(t)) 
 + \gamma_i\ {\bf K}_{0i}\ ({\bf x}_0 - {\bf x}_i)
\ \ \ \ \ \ \ \ i = 1, \ldots, n
\end{equation}
where ${\bf x}_0$ is the state of the leader, which we first
assume to be a constant. $\forall i$, ${\bf x}_i$
are the states of the followers; 
${\mathcal N}_i$ indicate the neighborship among the followers, where
time-delays are non-negligible in communications;
$\gamma_i=0$ or $1$ represent
the unidirectional links from the leader to the corresponding followers.
For each non-zero $\gamma_i$, the coupling gain ${\bf K}_{0i}$ is positive
definite.
\begin{theorem} \label{th:linear_schooling_leader_delayed}
Consider a leader-followers network~(\ref{eq:linear_schooling_leader_delayed}) 
with time-delayed communications. Regardless of the explicit values of the delays,
the whole system will tend to reach a group agreement
$$
{\bf x}_1(t) = \cdots = {\bf x}_n(t) = {\bf x}_0
$$ 
asymptotically if the whole network is connected, and the coupling links 
among the followers are either bidirectional with ${\bf K}_{ji} = {\bf K}_{ij}$, 
or unidirectional but formed in closed rings with identical gains.
\end{theorem}
{\bf Proof}: \ Exponential convergence of the leader-followers
network~(\ref{eq:linear_schooling_leader_delayed}) without delays has
been shown in~\cite{wei03-2, wei03-1} using contraction theory. If the
communication delays are non-negligible, and assuming that all the links
among the followers are bidirectional, we can transform the
equation~(\ref{eq:linear_schooling_leader_delayed}) to
$$
\dot{\bf x}_i\ =\ 
\sum_{j \in {\mathcal N}_i} {\bf G}_{ji} {\bf \tau}_{ji}
 + \gamma_i\ {\bf K}_{0i}\ ({\bf x}_0 - {\bf x}_i)
$$
where ${\bf \tau}_{ji}$ and correspondingly ${\bf \tau}_{ij}$ are defined
as the same as those in~(\ref{eq:network-wave-variable}). Considering the same
Lyapunov function $V$ as~(\ref{eq:delay-network-lyapunov}), we get
$$
\dot{V}\ =\ 
 -\ \sum_{i=1}^n \gamma_i \delta {\bf x}_i^T {\bf K}_{0i} \delta {\bf x}_i\
 -\ \frac{1}{2} \sum_{(i,j) \in {\mathcal N}} 
      ( \delta {\bf \tau}_{ji}^T \delta {\bf \tau}_{ji} + 
        \delta {\bf \tau}_{ij}^T \delta {\bf \tau}_{ij} )
$$
where ${\mathcal N}=\cup_{i=1}^n {\mathcal N}_i$ denotes the set of all
active links among the followers. Applying Barbalat's lemma shows that,
$\dot{V}$ will tend to zero asymptotically. It implies that,
$\forall i$ if $\gamma_i=1$, $\delta {\bf x}_i$ will tend to zero,
as well as $\delta {\bf \tau}_{ji}$ and
$\delta {\bf \tau}_{ij}$ $\forall (i,j) \in {\mathcal N}$.
Moreover, since
$$
\delta {\bf \tau}_{ji}(t)\ =\
{\bf G}_{ji}^T (\ \delta {\bf x}_j(t-T_{ji})\ -\ \delta {\bf x}_i(t)\ )
$$
we can conclude that, if the whole leader-followers network is
connected, the virtual dynamics will converge to
$$
\delta {\bf x}_1(t) = \cdots =  \delta {\bf x}_n(t) = 0
$$
regardless of the initial conditions or the delay values.
In other words, the whole system is asymptotically contracting.
All solutions will converge to a particular one, which in this case
is the point 
$
{\bf x}_1(t) = \cdots = {\bf x}_n(t) = {\bf x}_0
$. The proof is similar if there are unidirectional links formed in
closed rings. \hfill {\large \boldmath $\Box$}

\begin{figure}[h]
\begin{center}
\epsfig{figure=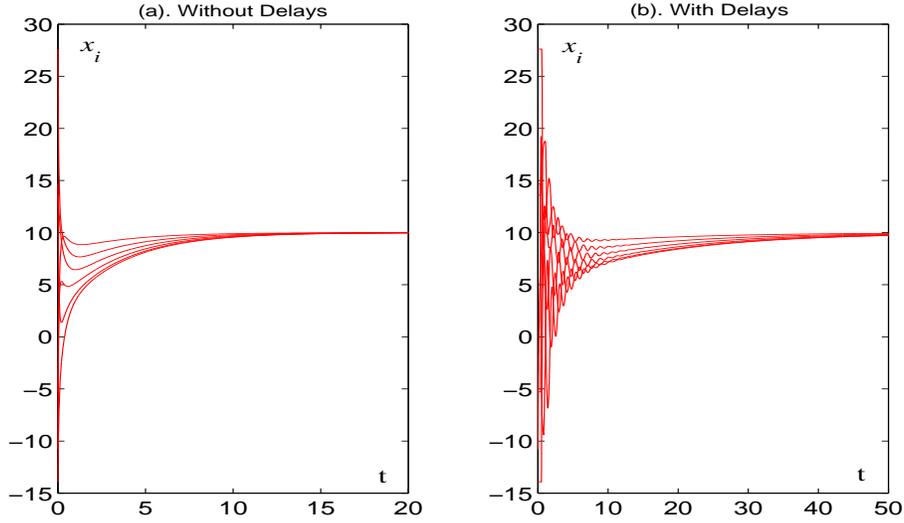,height=70mm,width=120mm}
\caption{Simulation results of Example~\ref{ex:flock-with-leader} without
delays and with delays. Initial conditions, chosen randomly, are
the same for each simulation.  In both cases, group agreement
to the leader value ${\bf x}_0$  is reached.}
\label{fig:delay-flock-with-leader}
\end{center}
\end{figure}

\Example{}{Consider a leader-followers
network~(\ref{eq:linear_schooling_leader_delayed}) with
one-dimensional ${\bf x}_i$, $n=6$, and structured as
$$ 
\xymatrix{ 0 \ar[r]  & 1 \ar[r] & 2 \ar[l] \ar[r] & 3 \ar[l] \ar[r] & 4 \ar[l] \ar[r] 
         & 5 \ar[l] \ar[r] & 6 \ar[l] } 
$$
The state of the leader is constant with value ${\bf x}_0=10$.
All the coupling gains are set to be identical with value $k=5$. The delay values
are not equal, each of which is chosen randomly around $0.5$ second.
Simulation results are plotted in Figure~\ref{fig:delay-flock-with-leader}.
}{flock-with-leader}

\begin{figure}[h]
\begin{center}
\epsfig{figure=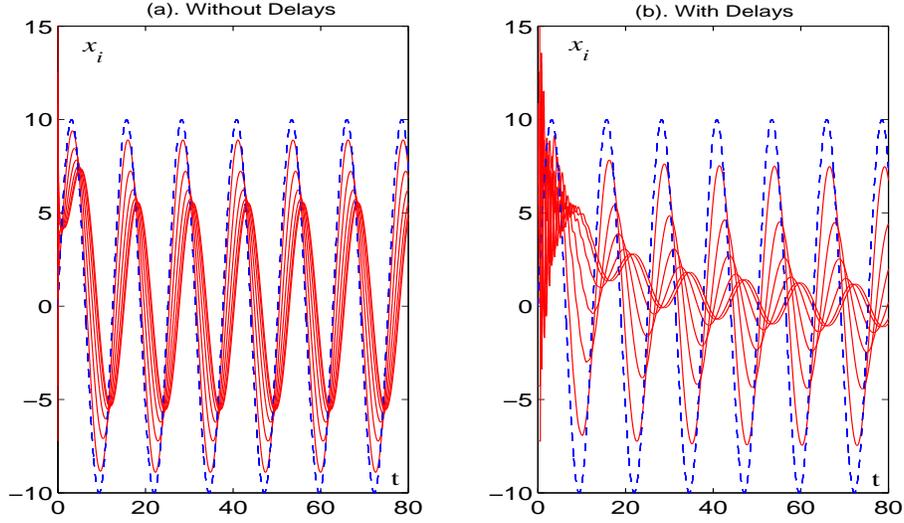,height=70mm,width=120mm}
\caption{Simulation results of Example~\ref{ex:flock-with-leader-varying} 
without delays and with delays. In both plots, the dashed curve is the state
of the leader while the solid ones are the states of the followers, which converge
to a periodic solution in both cases regardless of the initial conditions.}
\label{fig:delay-flock-with-leader-varying}
\end{center}
\end{figure}

Note that even if ${\bf x}_0$ is not a constant, i.e., the dynamics of
the $i$th element is given as
$$
\dot{\bf x}_i =  
\sum_{j \in {\mathcal N}_i} {\bf K}_{ji}\ 
          ({\bf x}_j(t-T_{ji}) - {\bf x}_i(t)) 
 + \gamma_i\ {\bf K}_{0i}\ ({\bf x}_0(t-T_{0i}) - {\bf x}_i(t))
$$
the whole system is still asymptotically
contracting according to exactly the same proof. Regardless of the initial
conditions, all solutions will converge
to a particular one, which in this case depends on the dynamics of ${\bf x}_0$ 
and the explicit values of the delays. Moreover, if ${\bf x}_0$ is periodic,
as one of the main properties of contraction~\cite{winni98}, 
all the followers' state ${\bf x}_i$ 
will tend to be periodic with the same period as ${\bf x}_0$.
\Example{}{
Consider the leader-followers network in Example~\ref{ex:flock-with-leader}
again. We set everything the same except that the leader's states is
not constant. Instead, we choose 
$$
{\bf x}_0=10 \ \sin (\frac{t}{2})
$$
The whole system is still asymptotically contracting regardless of the delays.
Simulation results are plotted in Figure~\ref{fig:delay-flock-with-leader-varying}.
}{flock-with-leader-varying}

\subsection{Discrete-Time Models} \label{sec:discrete-flocking-delayed}
Simplified wave variables can also be applied to study time-delayed
communications in discrete-time models.
Consider the model of flocking or schooling studied
in~\cite{jadbabaie03, viscek95}, where each element's state is 
updated according to the discrete-time law that computes the average of its 
neighbors' states plus its own state, i.e.,
\begin{equation} \label{eq:discrete_schooling}
x_i(t+1) \ =\ \frac{1}{1+n_i}\ (\ x_i(t)\ +\
\sum_{j \in {\mathcal N}_i} x_j(t)\ )
 \ \ \ \ \ \ \ \ \ \ \ \ \ \ \ \ \ \ \ \ \ i = 1, \ldots, n
\end{equation}
or, in an equivalent form
$$
x_i(t+1) \ =\ x_i(t)\ +\
\frac{1}{1+n_i}\ \sum_{j \in {\mathcal N}_i}\ 
(\ x_j(t)\ -\ x_i(t)\ )
$$
In fact, in equation~(\ref{eq:discrete_schooling}), there is 
no difference if $x_i$ is a scalar or a vector. For notation simplicity
we set $x_i$ as a scalar, but our analysis can be extended directly
to the vector case. In equation~(\ref{eq:discrete_schooling}), $\ t\ $ is
the index of the updating steps, so that its value is always an integer. 
$\ {\mathcal N}_i\ $ denotes the set of the active neighbors of
element $i$, which for instance can be defined as the set of the nearest 
neighbors within a certain distance $r$ around $i$. $\ n_i\ $
equals to the number of the neighbors of element $i$. As proved
in~\cite{jadbabaie03}, the whole system~(\ref{eq:discrete_schooling})
will tend to reach a group agreement if the network 
is connected in a very loose sense.

Assume now that time delays are non-negligible in communications.
The update law of the $i$th element changes to
\begin{equation} \label{eq:discrete_schooling_delayed}
x_i(t+1) \ =\ x_i(t)\ +\
\frac{1}{1+n_i}\ \sum_{j \in {\mathcal N}_i}\ 
(\ x_j(t-T_{ji})\ -\ x_i(t)\ )
\end{equation}
where the delay value $T_{ji}$ is an integer based on the number of 
updating steps. As in previous sections, $T_{ji}$ could be different
for different communication links, or even different along opposite 
directions on the same link.
For our later analysis, here we make a few assumptions:
all elements update their states synchronously, and the 
time interval between any two updating steps is a constant;
the network structure is fixed and always connected, which 
implies that $\forall i$, $n_i$ is a positive integer;
the value of neighborship radius $r$ is unique through the
whole network, which leads to the fact that all interactions are 
bidirectional.
\begin{theorem} \label{th:discrete_schooling_delayed}
Consider $n$ coupled elements~(\ref{eq:discrete_schooling_delayed}) with
time-delayed communications. Regardless of the explicit values of the delays,
the whole system will tend to reach a group agreement
$x_1(t) = \cdots = x_n(t)$ asymptotically.
\end{theorem}
{\bf Proof}: See Appendix. \hfill {\large \boldmath $\Box$}

Similarly, consider the discrete-time model of a cooperating group
with a leader-followers structure, where the dynamics of the $i$th
follower is given as
\begin{eqnarray*}
x_i(t+1) &=& \frac{1}{1+n_i+\gamma_i}\ (\ x_i(t)\ +\ 
\sum_{j \in {\mathcal N}_i}\ x_j(t-T_{ji})\ +\ \gamma_i x_0\ ) \\
&=& \frac{1+n_i}{1+n_i+\gamma_i} x_i(t)\ +\ 
\sum_{j \in {\mathcal N}_i}\ \tau_{ji}(t)\ +\
\frac{\gamma_i}{1+n_i+\gamma_i} x_0
\end{eqnarray*}
and $\tau_{ji}$ is defined as in
(\ref{eq:discrete-network-wave-variable}) with $k_i=1+n_i+\gamma_i$
and $\gamma_i=0$ or $1$. One easily shows that a very similar analysis
leads to the same result as that of
Theorem~\ref{th:linear_schooling_leader_delayed}.

\section{Concluding Remarks}
Modified wave variables are analyzed in this paper, and are shown to
yield effective tools for contraction analysis of interacting systems
with time-delayed feedback communications. Future work on time-delays
includes the development of analysis tools for general nonlinear
systems, coupled networks with switching topologies, and time-varying
time delays.


\section*{Appendix: Proof of Theorem~\ref{th:discrete_schooling_delayed}}

Equation~(\ref{eq:discrete_schooling_delayed}) can be transformed to
$$
x_i(t+1) \ =\ x_i(t)\ +\
\sum_{j \in {\mathcal N}_i} \tau_{ji}(t)
$$
where the wave variables are defined as
\begin{equation*}
u_{ji} = x_i \ + \ k_i \tau_{ji} \label{eq:discrete-network-wave-variable}
\ \ \ \ \ \ \ \ \ \ \ \ \ \ \ \ v_{ij} = x_i
\end{equation*}
and $k_i=1+n_i$.  Note that compared
to~(\ref{eq:network-wave-variable}), here we have ${\bf G}_{ij}={\bf
G}_{ji}=1$ (the cooperation law could be extended to a more general
form).  Define
$$
V(t)\ =\ \sum_{i=1}^n k_i \delta x_i^2(t)\ +\    
         \sum_{(i,j) \in {\mathcal N}}\ V_{i,j}(t)
$$
where $\forall (i,j) \in {\mathcal N}=\cup_{i=1}^n {\mathcal N}_i$
\begin{equation*}
V_{i,j}(t) \ = \ \sum_{\epsilon=t-1-T_{ij}}^{t-1} 
            \delta v_{ij}^2(\epsilon)\ +\
           \sum_{\epsilon=t-1-T_{ji}}^{t-1} 
            \delta v_{ji}^2(\epsilon) \ - \ \sum_{\epsilon=-T_{ij}}^0 
            \delta v_{ij}^2(\epsilon)\ -\
           \sum_{\epsilon=-T_{ji}}^0 
            \delta v_{ji}^2(\epsilon)
\end{equation*}
and therefore
\begin{eqnarray*}
V_{i,j}(t)
&=& \sum_{\epsilon=0}^{t-1}
     (\ \delta v_{ij}^2(\epsilon)\ -\
        \delta u_{ji}^2(\epsilon)\ +\
        \delta v_{ji}^2(\epsilon)\ -\
        \delta u_{ij}^2(\epsilon) \ ) \\
&=& -\ \sum_{\epsilon=0}^{t-1}
     (\ 2 k_i \delta x_i \delta \tau_{ji}(\epsilon)\ +\
        k_i^2 \delta \tau_{ji}^2(\epsilon)\ +\
        2 k_j \delta x_j \delta \tau_{ij}(\epsilon)\ +\
        k_j^2 \delta \tau_{ij}^2(\epsilon)\ )
\end{eqnarray*}
Since
\begin{eqnarray*}
\sum_{(i,j) \in {\mathcal N}} \big(V_{i,j}(t+1)\ -\ V_{i,j}(t)\big) 
&=& - \sum_{(i,j) \in {\mathcal N}}\ 
   (\ 2 k_i \delta x_i \delta \tau_{ji}(t)\ +\
        k_i^2 \delta \tau_{ji}^2(t)\ +\
        2 k_j \delta x_j \delta \tau_{ij}(t)\ +\
        k_j^2 \delta \tau_{ij}^2(t)\ ) \\
&=& - \sum_{i=1}^n\ (\ 
    2 k_i \sum_{j \in {\mathcal N}_i} \delta x_i \delta \tau_{ji}(t)\ +\
      k_i^2 \sum_{j \in {\mathcal N}_i} \delta \tau_{ji}^2(t)\ )
\end{eqnarray*}
one has
\begin{eqnarray*}
V(t+1) 
&=& \sum_{i=1}^n k_i (\ \delta x_i(t)\ +\ 
    \sum_{j \in {\mathcal N}_i} \delta \tau_{ji}(t)\ )^2\ +\    
    \sum_{(i,j) \in {\mathcal N}}\ V_{i,j}(t+1) \\
&=& V(t)\ +\ \sum_{i=1}^n k_i\ (\
    (\ \sum_{j \in {\mathcal N}_i} \delta \tau_{ji}(t)\ )^2\ -\
   k_i \sum_{j \in {\mathcal N}_i} \delta \tau_{ji}^2(t)\ )
\end{eqnarray*}
Note that
\begin{eqnarray*}
(\ \sum_{j \in {\mathcal N}_i} \delta \tau_{ji}\ )^2\ -\
   k_i \sum_{j \in {\mathcal N}_i} \delta \tau_{ji}^2 &=&
\delta \Gamma_i^T  
\left[ \begin{array}{ccc} 1 & \cdots & 1 \\ \vdots & \ddots & \vdots \\
                          1 & \cdots & 1 \end{array} \right]  
\delta \Gamma_i\ -\ k_i \delta \Gamma_i^T \delta \Gamma_i \\
&=& \delta \Gamma_i^T  
\left[ \begin{array}{cccc} -n_i & 1 & \cdots & 1 \\ 1 & -n_i & \cdots & 1 \\
    \vdots & \vdots & \ddots & \vdots \\ 1 & 1 & \cdots & -n_i \end{array} \right]  
\delta \Gamma_i \\
&\le&-\ \delta \Gamma_i^T \delta \Gamma_i\ =\ -\
\sum_{j \in {\mathcal N}_i} \delta \tau_{ji}^2
\end{eqnarray*}
where $\ \delta \Gamma_i=[\cdots,\ \delta \tau_{ji},\ \cdots]^T \in \mathbb{R}^{n_i}\ $, so that
$$
V(t+1)\ \le\ V(t)\ -\ 
\sum_{i=1}^n k_i\ \sum_{j \in {\mathcal N}_i} \delta \tau_{ji}^2(t)
$$ 
Since $V$ is lower bounded it converges to a finite limit as $t
\rightarrow +\infty$, which implies that $\forall (i,j) \in {\mathcal
N}$, $\delta \tau_{ji}$ tends to zero, which in turn implies that
$\forall i$, $\delta x_i$ tends to zero.  The rest of the proof is
then similar to that of Theorem~\ref{th:linear_schooling_delayed}.

%
%
\renewcommand{\baselinestretch}{1.}

\end{document}